\UseRawInputEncoding
\documentclass[twocolumn,prl,amsmath,amssymb,showpacs,superscriptaddress,floatfix]{revtex4}
\usepackage{graphicx}% Include figure files
\usepackage{bm}% bold math
\usepackage{lineno,hyperref}
\usepackage{epstopdf}
\usepackage{epsf}
\usepackage{xcolor}
\begin{document}
%\draft
\title{First-principles molecular dynamics simulation of liquid indium}%:
%First-Order versus Continuous Transition }

\author{Yu. D. Fomin \footnote{Corresponding author: fomin314@mail.ru}}
\affiliation{Vereshchagin Institute of High Pressure Physics, Russian Academy of Sciences,
Kaluzhskoe shosse, 14, Troitsk, Moscow, 108840 Russia}
\affiliation{Moscow Institute of Physics and Technology (National Research University), 9 Institutskiy Lane, Dolgoprudny, Moscow region, 141701, Russia}

\date{\today}

\begin{abstract}
We report an ab-initio simulation of liquid Indium in a wide range of pressures and temperatures. We calculate equation of state, thermal expansion and compressibility coefficients. The structure of the system is analyzed by radial distribution functions and structure factors. The results are compared with available experimental data.

\end{abstract}


\maketitle

\section{Introduction}

%Thermodynamic, structural and dynamical properties of liquids %in a wide range of pressures are of great importance for many %fundamental and applied problems. 

Liquid indium is an important metal in many fields of industry. It can be used as coverege to reduce friction coefficient, as a solder, as an additive to glasses to modify their optical properties, etc. It also attracted attention of researchers for many years. Since 196-th there were several important works which studied the structure of liquid indium at different temperatures \cite{str-1,str-2,str-3}. Most of them, however, considered the system at ambient pressure only. At the same time the melting curve of indium is measured up to rather high pressure of 10.5 GPa \cite{dudley} (see also \cite{mc} for the melting curve up to about 1 GPa).

Basing on the results of Refs. \cite{dudley} and \cite{mc} the authors of Ref. \cite{shen} studied the structure and the density of liquid indium up to the melting line at $T=710$ K. Basing on X-ray techniques the authors measured the density of molten indium at several pressures up to the melting point and the structure factors. They also determined the coordination number by integrating of radial distribution functions $g(r)$. From their results one can see that the coordination number monotonically increases with pressure. 

Another X-ray study of liquid indium is reported in Ref. \cite{mudry}. In this study experiments were performed at ambient pressure and a set of temperatures. The authors report that the coordination nember behaves non-monotonically and conclude that a liquid-liquid phase transition takes place. 

In Ref. \cite{shen1} the molar volume of liquid indium at $T=710$ K was measured up to $P=8.5$ GPa. Later on these data were fitted to Birch-Murnaghan equation in Ref. \cite{liq-in}. Basing on these fits the authors calculated the bulk modulus and thermal expansion coefficient of liquid indium along the $T=710$ K isotherm.

In the present work we perform ab-initio molecular dynamics calculations of liquid indium. We calculate the equation of state of the system, its bulk modulus and thermal expansion coefficient. We also compute the structure factors. All quantities are compared with the experimental data where it is possible.

\section{System and methods}

In the present study we investigate a system of 100 atoms of indium in a cubic box with periodic boundary conditions. The system is simulated by means of ab-inition molecular dynamics method as realized in VASP simulation package. Projector augmented wave method is used for the treatment of the electronic structure. The energy cut-off is set to 96 eV. The time step is set to 3 fs. Only gamma point was taken into account in the ${\bf k}$ space calculations.

\begin{figure}
\includegraphics[width=8cm,height=8cm]{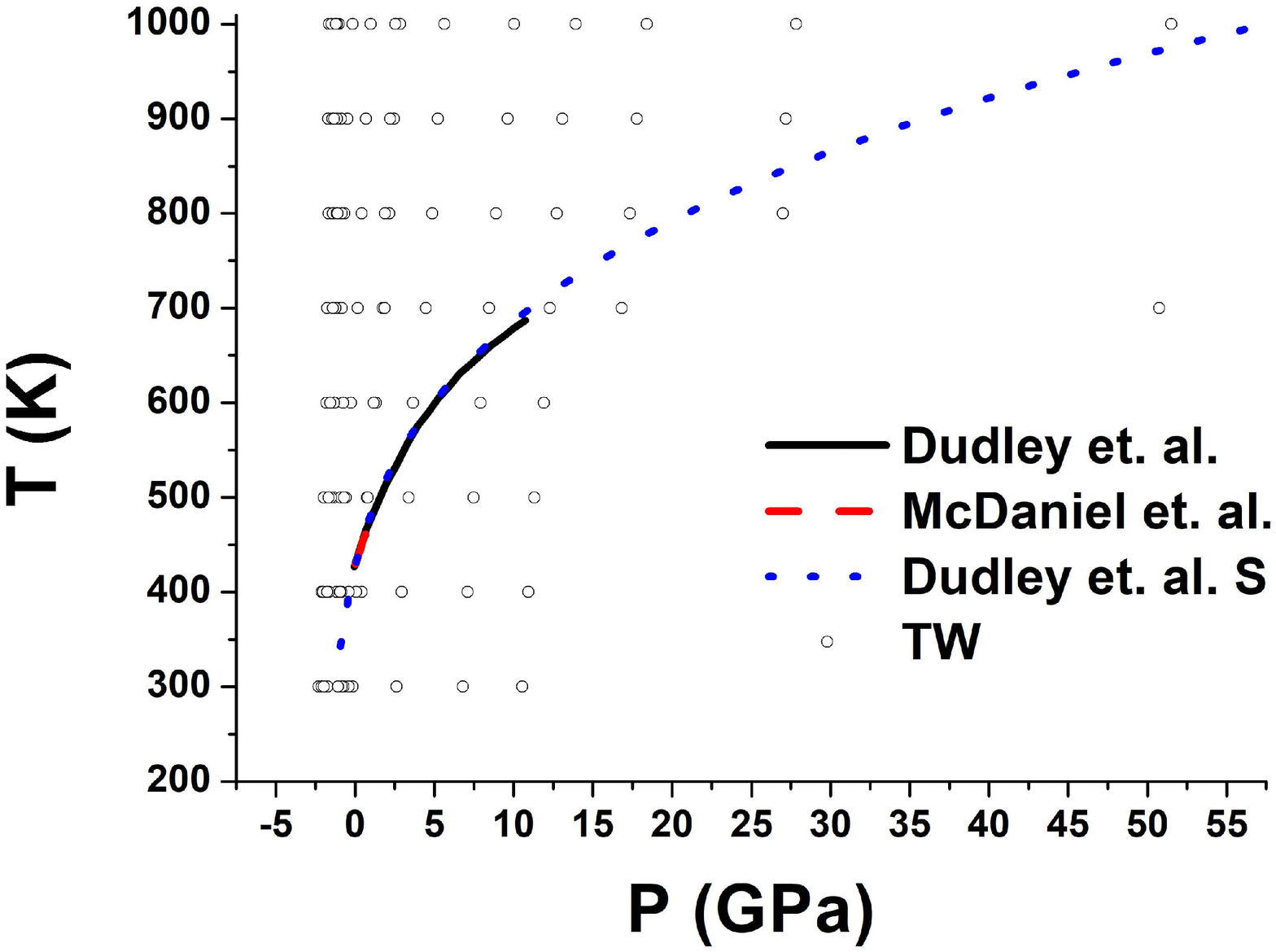}%

\caption{\label{pd-exp-md} Points where simulations were performed on the phase diagram. The curve 'Dudley et. al.' shows the results from Ref. \cite{dudley}, the curve 'McDaniel' - results from Ref. \cite{mc}, 'Dudley et. al. S' - fitting the results of Ref. \cite{dudley} to Simon equation (Eq. (2) of Ref. \cite{dudley}), TW - the points of the present work.}
\end{figure}

Large set of densities and temperatures is considered. The densities varied from $\rho_{min}=3.88$ to $\rho_{max}=19.06$ $g/cm^3$. At each density we firstly simulated the high temperature system with $T_{max}=1000$ K. Then the last configuration of this simulation was used as an initial one for simulation of the system at the temperatures from $T=300$ up to $T=900$ K with the step of $100$ K. Each simulation lasted for 10000 steps, i.e. 30 ps. The set of simulated points is shown on the phase diagram in Fig. \ref{pd-exp-md}

\section{Results and discussion}

At the first step we calculate the equation of state on liquid indium and compare it to the experimental data from ref. \cite{liq-in}.

\begin{figure}
\includegraphics[width=8cm,height=8cm]{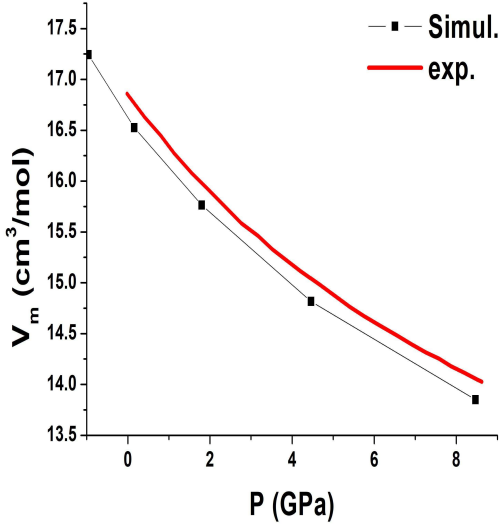}%

\caption{\label{md-exp} Comparison of experimental data at $T=710$ K from Ref. \cite{liq-in} and the data from ab-initio simulation of this work at $T=700$ K.}
\end{figure}

Fig. \ref{md-exp} presents a comparison of the data of this work and the experimental results from \cite{liq-in}. The experimental data are taken at $710$ K, while in the present work $T=700$ K is used. We believe that this difference sufficiently small to not affect the comparison. 

The difference between the experimental and simulation data does not exceed $1.12 \%$. Basing on this we conclude that the results of simulations are reliable.

In order to evaluate the equation of state of liquid indium on more solid basis we fit it to a polynomial function 

\begin{equation}
P(\rho,T)=\sum_{n,m}a_{n,m}\rho^{n}T^m,
\end{equation}
where $\rho$ is density in $g/cm^3$, T is temperature in Kelvins, $a_{n,m}$ are fitting coefficients and the exponents n and m are choosen in such a way that $n+m <=5$.

%Analogosly we fit the internal energy of the system to the %equation

%\begin{equation}
%E(\rho,T)=\sum_{n,m}b_{n,m}\rho^{n}T^m,
%\end{equation}\label{fit-en}
%where $E$ is energy in eV and $b_{n,m}$ are fitting %coefficients. 

\begin{figure}
\includegraphics[width=8cm,height=8cm]{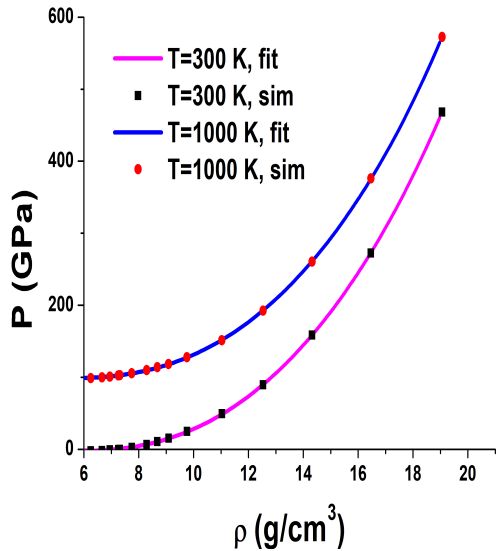}%

\caption{ Comparison of the data from ab-inition molecular dymanics and fitting to Eq. \ref{fit-pr}. The data for $T=1000$ K are shifted to 100 GPa upwards in order to avoid overlaps.}\label{fit}
\end{figure}

Fig. \ref{fit} shows a comparison of the data for equation of state from ab-inition simulation and from fitting to Eq. (1). One can see that the quality of the fitting is perfect. Below we use this fitted equation of state for calculation of the response functions $B_T=\rho \left( \frac{\partial P}{\partial \rho} \right)_T$ and $\alpha_P=-\frac{1}{T} \left( \frac{\partial \rho}{\partial T} \right)_T$. Similarly, fitting of the energy is used for calculation of the heat capacity of the system.

\begin{table}
\begin{tabular}{ l | c | r }
  \hline			
  n & m & $a_{n,m}$ \\
  \hline			
  0 & 0 & -305.57333657145500 \\
  1 & 0 & 238.70598753111230 \\
  2 & 0 & -63.833351593789722 \\
  3 & 0 & 6.5433130551773946                 \\
  4 & 0 &  -0.24008639734032150 \\
  5 & 0 &  3.9155839840238878E-003\\
  0 & 1 &   -0.17373675573617220\\
  1 & 1 &  1.4237276981756433E-002\\
  2 & 1 & 8.0437749691983397E-003 \\
  3 & 1 &    -5.7392554358567341E-004\\
  4 & 1 &   1.5599905130353485E-005 \\
  2 & 2 &  5.8408548738952959E-004 \\
  1 & 2 &    -1.3516570420262682E-004 \\
  2 & 2 & -1.5393806829344470E-006 \\
  3 & 2 & -3.4804293612014242E-008 \\
  0 & 3 & -5.1700880554506057E-007 \\
  1 & 3 & 1.7197816344291688E-007 \\
  2 & 3 & 1.3138966034853920E-009 \\
  0 & 4 & -6.8034449601794833E-011 \\
  1 & 4 & -7.7188558042335661E-011 \\
  0 & 5 & 1.8898204983398251E-013 \\

  \hline  
\end{tabular}
\caption{Coefficients of Eq. (1). The resulting pressure is in kbars (0.1 GPa). The dimensionality of coefficients $a_{n,m}$ are choosen in such a way that $a_{n,m}\rho^{n}T^m$ is in kbars. }
\end{table}

\begin{figure}
\includegraphics[width=8cm,height=8cm]{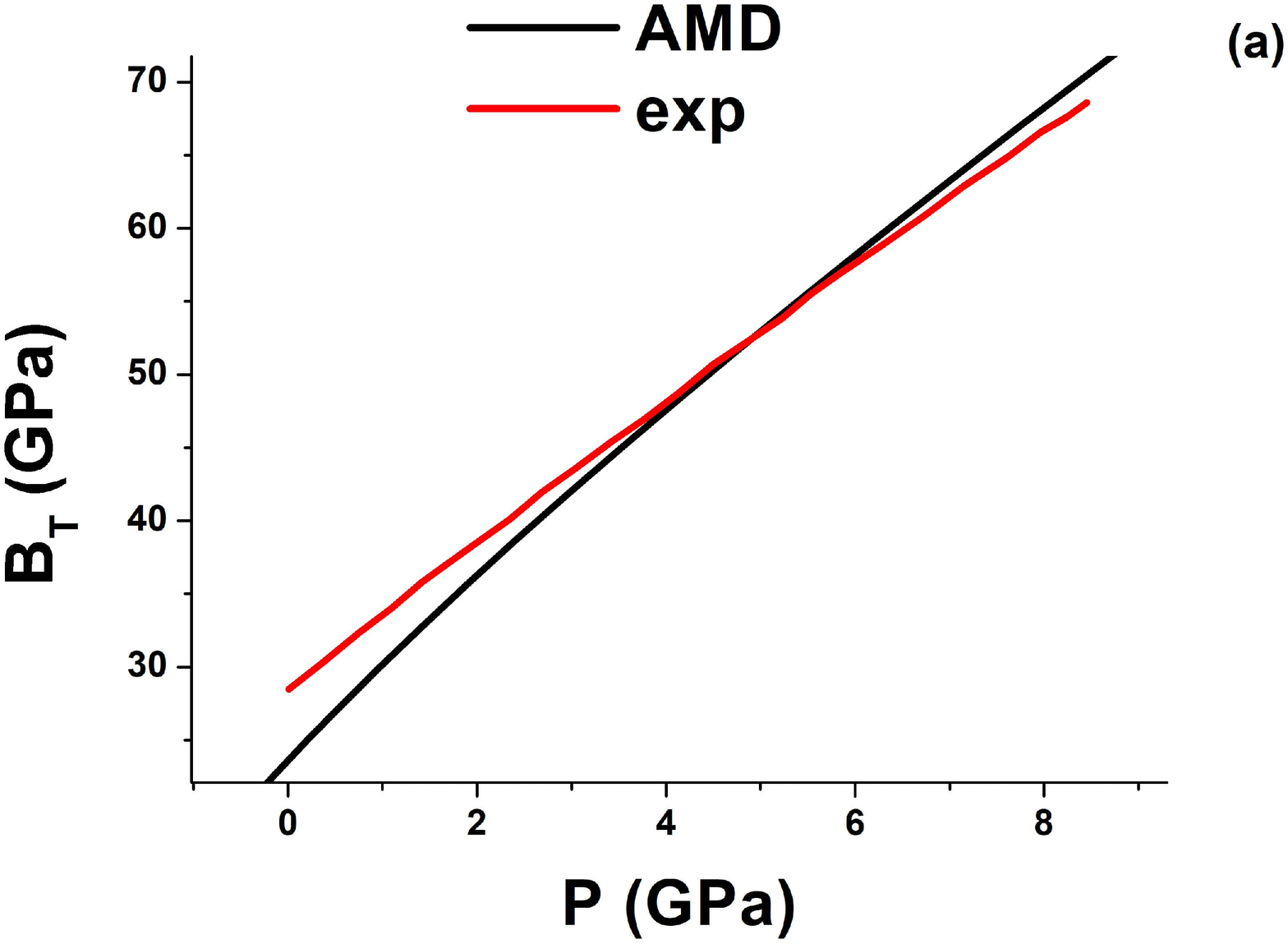}%

\includegraphics[width=8cm,height=8cm]{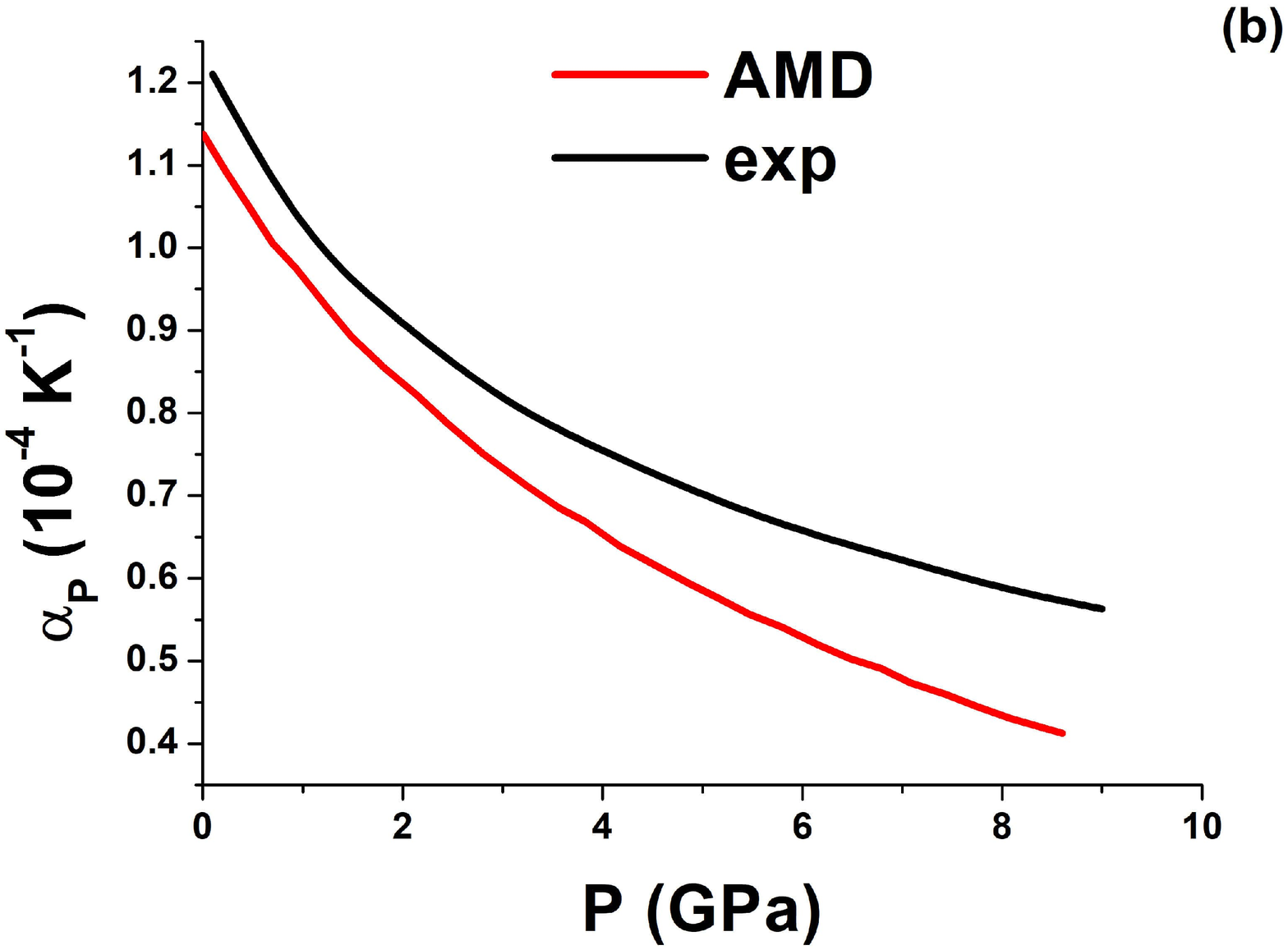}%

\caption{(a) Comparison of bulk modulus $B_T$ at $T=700$ K from ab-initio simulation of the present work (AMD) and from experiments of Ref. \cite{liq-in}. (b) The same for the thermal expansion coefficient. (c) The same for isochoric heat capacity $c_V$ (in units of $k_B$). The straight line shows $c_V=2k_B$ level.}\label{btal}
\end{figure}

Fig. \ref{btal} (a) and (b) show a comparison of the data of the present calculations with the experimental data from Ref. \cite{liq-in} for the bulk modulus and thermal expansion coefficient. In the case of bulk modulus the curves are very close. However, the slope of the calculated curve is slightly larger. The largest descreepancy takes place at low pressure and reaches about $20 \%$.

The agreement between simulation and experement becomes worse in the case of thermal expansion coefficient. Experemental values are smaller then the calculated ones. The descreepancy increases with pressure reaching about $40 \%$ at $P=8$ GPa.

%Fig. \ref{btal} (c) shows the isochoric heat capacity of the %system along the same isotherm $T=700$ K. 

\begin{figure}
\includegraphics[width=8cm,height=8cm]{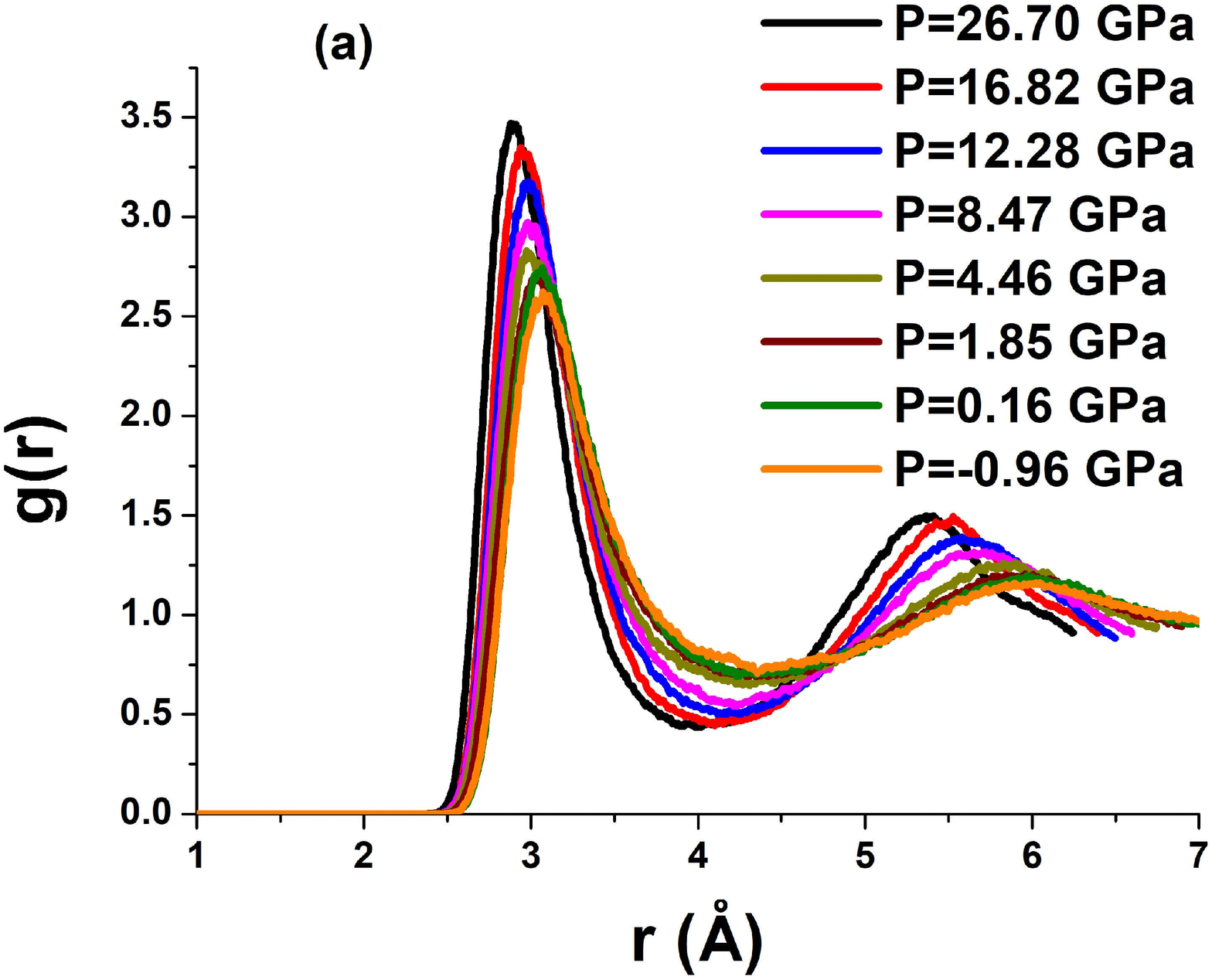}%

\includegraphics[width=8cm,height=8cm]{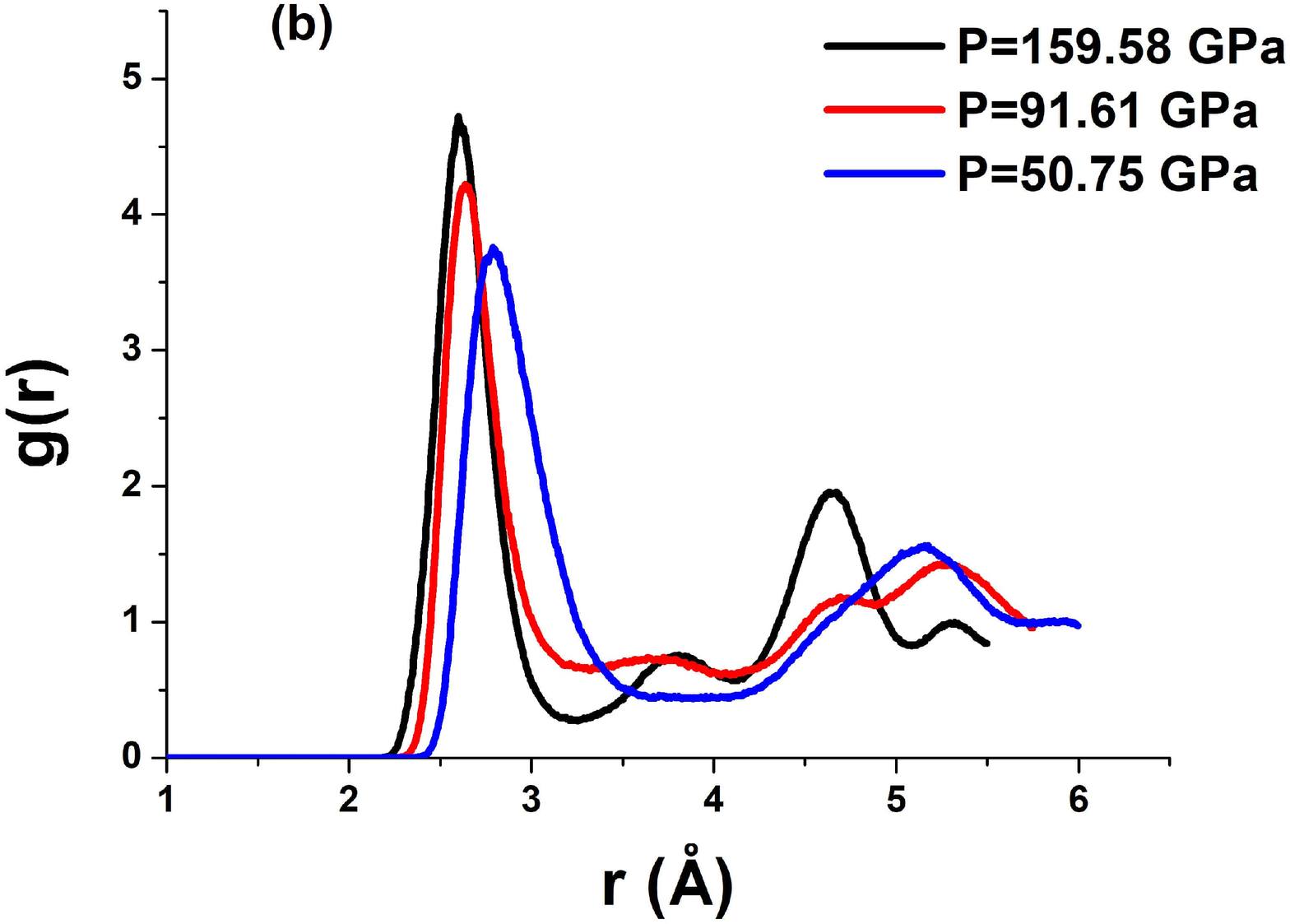}%

\caption{(a) Radial distribution functions of liquid indium at $T=700$ K isotherm for the pressures from $P=-0.96$ GPa up to 26.60 GPa. (b) The same for higher pressures.}\label{rdf}
\end{figure}

In order to characterize the structure of the liquid we calculated the radial distribution functions $g(r)$. Fig. \ref{rdf} shows $g(r)$ for a set of pressures along the isotherm $T=700$ K. According to Ref. \cite{liq-in} the melting point of indium at $T=710$ K is $P_m=7.77$ GPa. From Fig. \ref{rdf} (a) one can see that even at the pressure as high as $P=26.70$ GPa the system preserves liquid-like structure which should be related to formation of metastable liquid in ab-initio simulations. Some signs of crystallinity can be observed at pressures as high as $P=50.75$ GPa whish is much higher then experimental melting pressure.

\begin{figure}
\includegraphics[width=8cm,height=8cm]{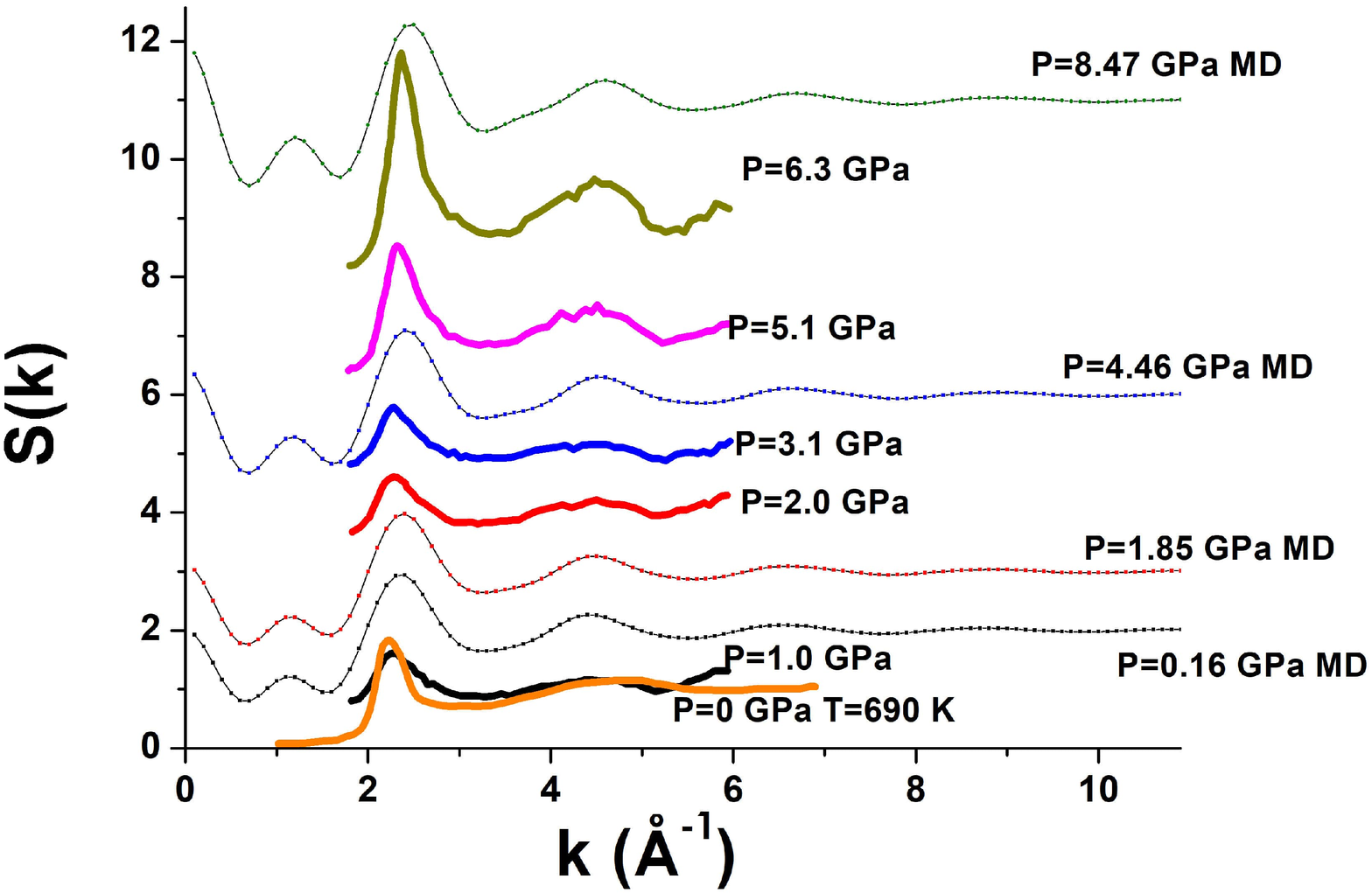}%

\caption{Comparison of the structure factors obtained in the present work with the experimental ones. The curves are shifted upward to avoid overlaps. Pressures are given next to the curves. The curve at $P=0$ GPa and $T=690$ K is from Ref. \cite{mudry}. The curves at $P=1.0$, $2.0$, $3.1$, $5.1$ and $6.3$ GPa are from Ref. \cite{shen}. The curves with label 'MD' after the pressure are our calculations.}\label{sfac-exp}
\end{figure}

X-ray experiments measure structure factor $S({\bf k})$ which is a Fourier transform of radial distribution function. Fig. \ref{sfac-exp} shows the structure factor of the system obtained in the present work in comparison to the experimental ones from Refs. \cite{shen,mudry}. The structure factors of this work are obtained by numerical Fourier transform of $g(r)$. The results for $k<1.6$ $\AA^{-1}$ should be addressed to numerical errors and should be discarded from consideration. 

One can see that experimental structure factors demonstrate very slow decay of the right branch of the first peak. Unlike this the first peak of $S({\bf k})$ of our calculations look more symmetric and 'Gauss-like'. However, the overall aggreement is sufficiently good.

\begin{figure}
\includegraphics[width=8cm,height=8cm]{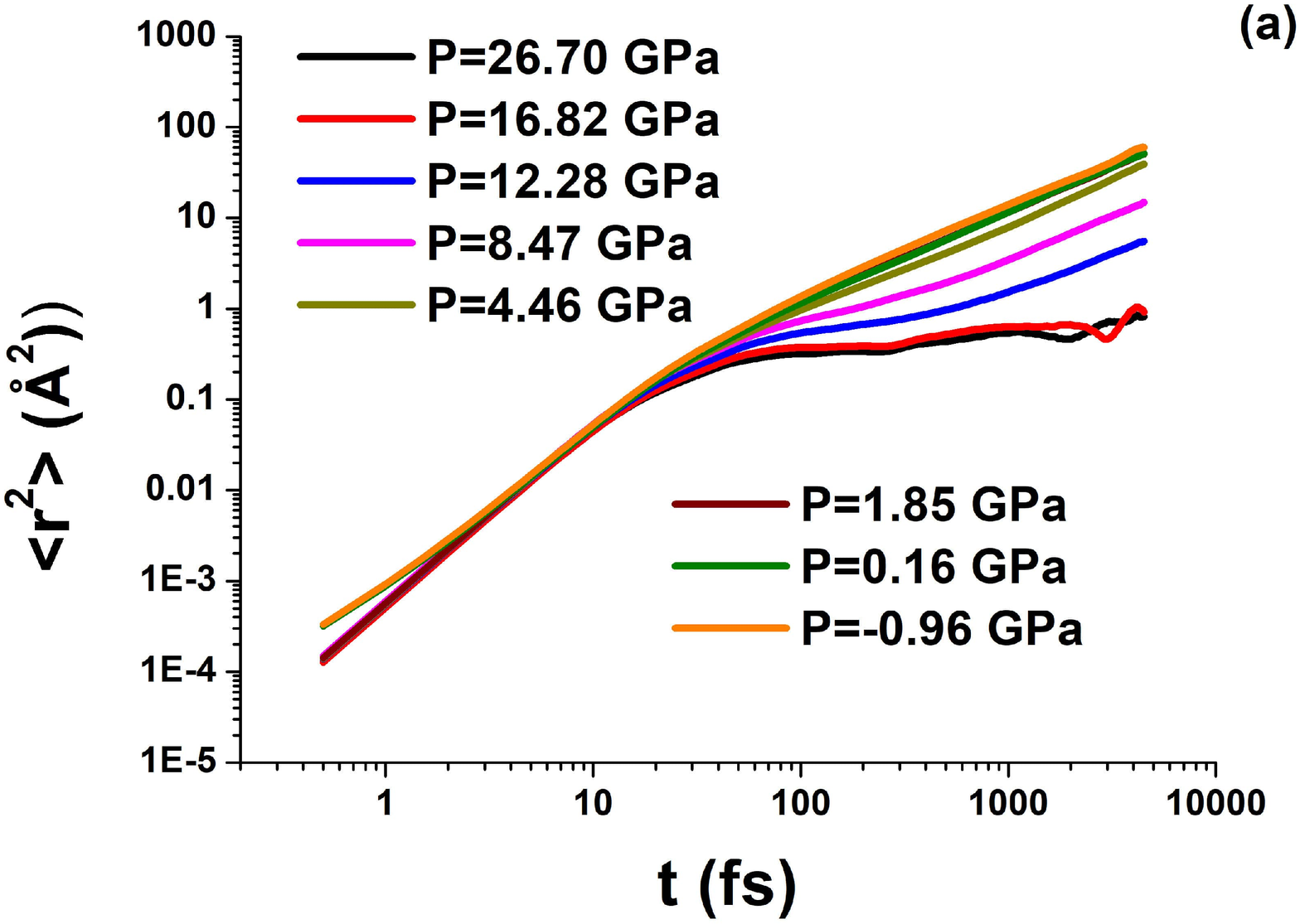}%

\includegraphics[width=8cm,height=8cm]{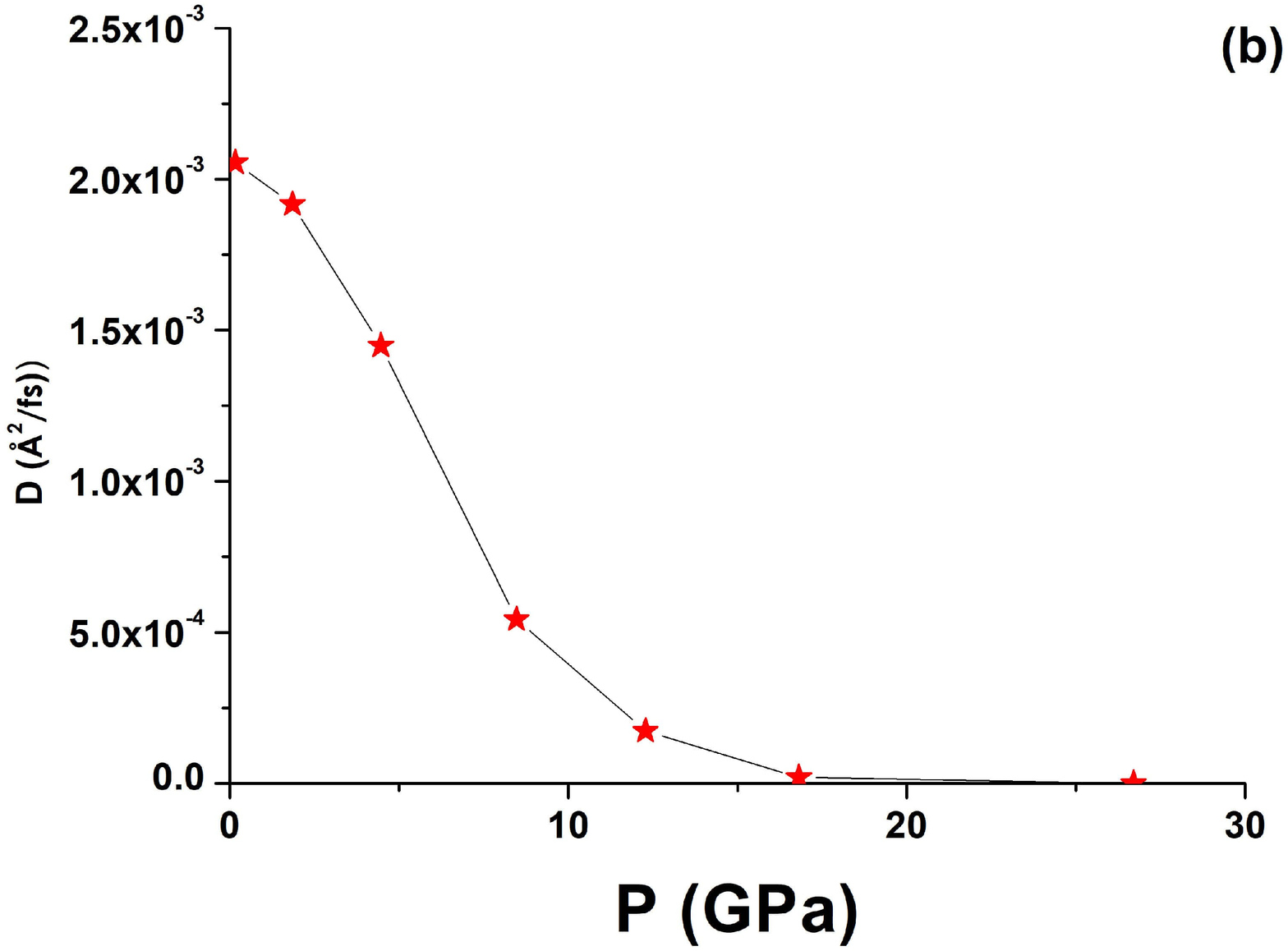}%

\caption{(a) Mean square displacement of liquid indium at a set of pressures at $T=700$ K. (b) Diffusion coefficient of liquid indium along the same isotherm.}\label{msd}
\end{figure}

Although the crystallization is not visible in radial distribution functions, it is more pronounced in the dynamical properties of the system characterized by means square displacement. Fig. \ref{msd} (a) shows mean square displacement long the $T=700$ K isotherm. One can see that some mobility of atoms preserves up to the pressure $=12.28$ GPa and disappearsh at higher pressures. Fig. \ref{msd} (b) demonstrates the diffusion coefficient along the same isotherm. One can see that it monotonously decreases with pressure and vanishes at $P=16.82$ GPa.

This work was carried out using computing resources of the federal collective usage center "Complex for simulation and data processing for mega-science facilities" at NRC "Kurchatov
Institute", http://ckp.nrcki.ru, and supercomputers at Joint
Supercomputer Center of the Russian Academy of Sciences (JSCC
RAS). The work was supported by the Council of the President of the Russian Federation for State Support of Young Scientist (Grant MD-6103.2021.1.2).

\end{document}